# The role of Jupiter in driving Earth's orbital evolution

Jonathan Horner [1,2], Dave Waltham [3] and F. Elliot Koch [1,2,4]

[1] *School of Physics, University of New South Wales, Sydney, NSW 2052, Australia*
[2] *Australian Centre for Astrobiology, University of New South Wales, Sydney, NSW 2052, Australia*
[3] *Department of Earth Sciences, Royal Holloway, University of London*
[4] *San Diego State University, Physics Department, San Diego, CA 92182-1233, USA, 5500 Campanile Drive*

**Summary:** In coming years, the first truly Earth-like planets will be discovered orbiting other stars, and the search for signs of life on these worlds will begin. However, such observations will be hugely time-consuming and costly, and so it will be important to determine which of those planets represent the best prospects for life elsewhere. One of the key factors in such a decision will be the climate variability of the planet in question - too chaotic a climate might render a planet less promising as a target for our initial search for life elsewhere.

On the Earth, the climate of the last few million years has been dominated by a series of glacial and interglacial periods, driven by periodic variations in the Earth's orbital elements and axial tilt. These Milankovitch cycles are driven by the gravitational influence of the other planets, and as such are strongly dependent on the architecture of the Solar system.

Here, we present the first results of a study investigating the influence of the orbit of Jupiter on the Milankovitch cycles at Earth - a first step in developing a means to characterise the nature of periodic climate change on planets beyond our Solar system.

## Introduction

Ever since we first looked at the night sky, mankind has asked the question "Are we alone?". Until just two decades ago, the answer to that question was solely the preserve of fiction, fantasy, speculation and faith – we knew of no planets beyond our Solar system. As such, it was an open question as to whether our Solar system was a fluke, miraculously alone in the universe, or whether planets were instead common around other stars.

The discovery of the first planets orbiting other Sun-like stars (e.g. [1][2][3]) was a watershed moment. Over the years, an ever-growing catalogue of exoplanets have been discovered[1], and have revealed that the variety of planets orbiting other stars is far more diverse than we could ever have imagined – ranging from Jupiter and Earth-sized planets that orbit within a million kilometres of their host star's surface (e.g. [4][5][6]), to planets orbiting binary stars (e.g. [7][8]), proposed "diamond planets" (e.g. [9][10]), and even a few moving on highly eccentric orbits reminiscent of the comets in our own Solar system (e.g. [11]). Coupled with our ever increasing understanding of our own Solar system (e.g. [12][13][14]), these discoveries have revolutionised our understanding of the formation and evolution of planetary systems (e.g. [15][16][17]).

---

[1] For an up-to-date tally of the number of known exoplanets, we direct the interested reader to the two main catalogues of exoplanets on the internet – http://exoplanets.org/ and http://exoplanet.eu/ . As of 1st November, 2013, these websites give the current tally of "confirmed" exoplanets as 755 and 1038, respectively.

In the coming years, it is highly likely that we will discover the first truly Earth-like planets[2] orbiting nearby stars, and the search for life beyond our Solar system will be able to begin in earnest. However, the observations required to detect evidence of life on Earth-like planets orbiting other stars will be hugely time-consuming and costly – which will in turn mean that we will only be able to focus on the few most promising targets, at least in the early part of that search. So how will we chose which of those exo-Earths we should target in the search for life? Clearly, the proximity of the various exo-Earths to our Solar system will be an important factor in any decision – the closer an exoplanet to Earth, the more widely it will be separated from its host star, and the easier it will therefore be to observe. There are, however, a variety of additional factors that will play an important role in determining which of the exo-Earths we discover are the most promising targets for the search for life (e.g. [18][19].

One factor which will play an important role in any decision on exo-Earth suitability is the climatic stability of that planet. It is reasonable to assume that life will have the greatest chance of becoming established and thriving on a planet with a relatively quiescent climate. In contrast, planets whose climates oscillate dramatically and rapidly will likely represent poor targets for our initial search for life[3]. It is well established that the Earth's climate varies on timescales of tens and hundreds of thousands of years as a result of variations in our planet's orbit driven by the gravitational influence of the other planets in our Solar system. The Earth's orbit twists and flexes as it is pulled around by the other massive objects in the Solar system, which results in long-term variations in the amount of energy the Earth receives from the Sun, averaged over any given year. These oscillations, known as the Milankovitch cycles, were first recognised in the late 19th and early 20th Century (e.g. [20][21][22]), and are thought to have driven the recent series of glacial and interglacial periods that have occurred over the last few million years (e.g. [23][24]).

Beyond the most recent ice age, there is evidence that the Earth has experienced at least four other periods of glaciation through its history. The Huronian glaciation, over two billion years ago, is thought to have lasted approximately 300 Myr, and may well have been linked to the evolution of photosynthesis (which would have removed a significant amount of greenhouse gas from the atmosphere, causing the planet to cool) [25]. More recently, the Sturtian and Marinoan glaciations (~750-700 Myr and 635 Myr ago; [26][27]) are thought to have been some of the most extensive glaciations in our planet's history – the archetypal "snowball Earth" events. It is almost certain that, during these periods of global glaciation, the Milankovitch cycles will have played an important role in driving the backward and forward march of the ice-caps on timescales of tens and hundreds of thousands of years. Indeed, studies of the glaciation that occurred at the end of the Ordovician and in the early Silurian periods (c.a. 445 Myr ago; [28]) have shown strong evidence of cyclical variations that can best be explained by the Milankovitch cycles of the time (e.g. [29]).

Whilst for the Earth, the periodic orbital variations induced by the other planets are relatively small, the same cannot be said for the planet Mercury. At the current epoch, Mercury moves on an orbit with eccentricity approximately 0.21. On timescales of hundreds of thousands of years, Mercury's orbital eccentricity varies dramatically – at times falling so low as to put the planet on a circular orbit, whilst at other times being driven as high as 0.45 (e.g. [30][31]). Were the Earth's orbital excursions equally extreme, there would be times when our planet's orbit would take it significantly closer to the Sun than the planet Venus, at perihelion, and out almost to the orbit of Mars, at aphelion. Such a situation would clearly have deleterious effects on our planet's climate, and would likely render it a less hospitable place for the development of life.

---

[2] i.e. planets of comparable size and mass to the Earth, orbiting at a suitable distance from their host star such that liquid water could be present and stable upon their surface.
[3] Although recent work by Spiegel et al. (e.g. [45]) has emphasised that such behaviour could also be advantageous to a planet's habitability, in particular since it could aid the rapid emergence of that planet from snowball-Earth states.

In this work, we present the preliminary results of a study that aims to categorise the influence of the giant planet Jupiter's orbit on the scale and speed of the Milankovitch cycles experienced by the Earth. The key goal of this work is to build expertise that will allow future studies to determine the Milankovitch cycles that would be experienced by any newly discovered exo-Earths, in order to help determine which would be the most promising targets for the search for life. In section two, we describe our methodology, before presenting our preliminary results in section three. Finally, in section four, we conclude with a discussion of the future direction of our research project.

## Testing Jupiter's Role

To examine the influence of Jupiter's orbit on the amplitude and frequency of the Milankovitch cycles of the Earth, we used the Hybrid integrator within the *n*-body dynamics package MERCURY [32], which has been widely used to address questions in Solar system astronomy, exoplanetary science and astrobiology (e.g. [33][34][35]). In order to properly account for the effects of general relativity on the orbit of the planet Mercury, we used a version of the MERCURY code that had been modified to take account of the relativistic correction to Mercury's orbital motion. We then set up a total of 39,601 unique simulations, each of which ran for a simulation period of one million years, from the current epoch forward in time. In each simulation, the initial orbits of Mercury, Venus, Earth, Mars, Saturn, Uranus and Neptune were held constant at their current values. The initial orbit of Jupiter, however, was systematically varied in semi-major axis and eccentricity. In total, we tested 199 unique values of the initial Jovian semi-major axis, ranging from 4.20336301 to 6.20336301 AU, in equal steps (i.e. covering a range ±1 AU from the orbit of Jupiter in our Solar system, at *a* = 5.20336301 AU). At each of these Jovian semi-major axis values, we test 199 unique orbital eccentricities, distributed evenly between 0.0 and 0.2 (for reference, the current value of Jupiter's orbital eccentricity is 0.04839266)[4]. The orbital elements for the eight planets were recorded at 100 year intervals for the duration of the 1 Myr integrations.

Once the simulations were complete, we used the results to create maps of the variability of the Earth's orbital elements as a function of Jupiter's initial orbit, building on earlier work creating dynamical maps of exoplanetary systems and the orbits of Solar system objects (e.g. [36][37]). These maps provide a quick visual guide to the degree of variability in the Earth's Milankovitch cycles that can result from small scale changes to the orbit of Jupiter, and we present a number of examples of such plots in the next section.

## Preliminary Results

In Figure 1, we present four exemplar plots that reveal how moderate changes to Jupiter's orbit can have significant effects on both the periodicity and the amplitude of the

---
[4] The orbital elements for the eight planets were taken from
http://www.met.rdg.ac.uk/~ross/Astronomy/Planets.html, and are valid for epoch JD 2451545.0 (i.e. January 1.5, 2000, UT).

Milankovitch cycles experienced by the Earth. In that figure, the left hand panels show the evolution of the Earth's orbital eccentricity (top) and inclination (bottom) for a period of one million years in a Solar system where Jupiter occupies its current orbit around the Sun. The right hand panels show the variation of the same variables for a Solar system that differs from the first only in the initial eccentricity of Jupiter's orbit. In the right hand panels, rather than Jupiter starting on the low eccentricity orbit we see in our Solar system ($e \approx 0.048$), the giant planet instead moved on an orbit with an initial eccentricity 0.2 (still far less than the eccentricities proposed for a number of candidate exoplanets; e.g. [38][39]). As can be seen, the variations in Earth's orbital eccentricity differ greatly in scale between the two scenarios (with the high eccentricity Jupiter driving oscillations approximately five times larger than those in our Solar system). On the other hand, inclination variations are comparable in scale between the two scenarios, but the more eccentric Jupiter forces the Earth's orbital inclination to vary at a higher frequency than observed in the Solar system. Whilst these are just the results of two simple simulations, they serve to illustrate the effect that variations in the orbits of the other planets can have on the Milankovitch cycles at Earth.

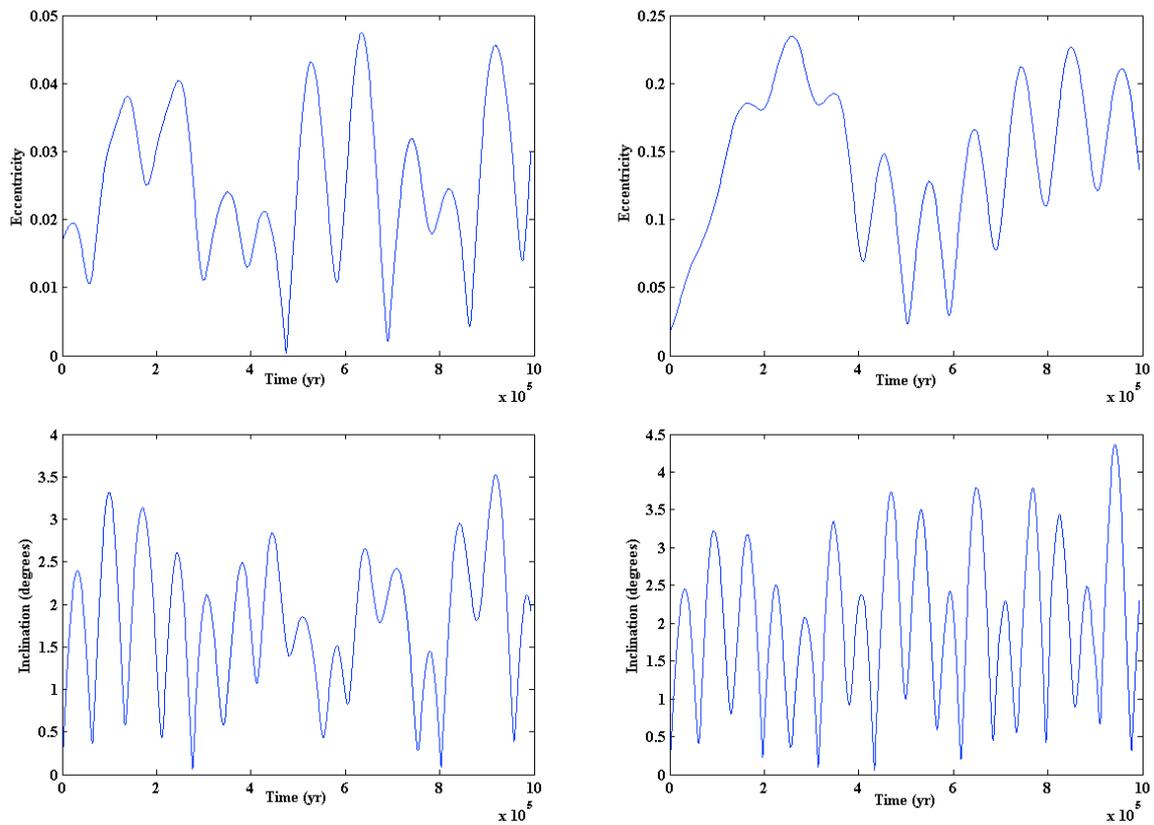

*Fig. 1: The evolution of the eccentricity (top) and inclination (bottom) of the Earth's orbit, for a period of one million years. The left hand plots show the variations in eccentricity and inclination for the Solar system as we know it today, with Jupiter on its current orbit. Those to the right show the variability if Jupiter instead moved on an initial orbit with eccentricity 0.2, but with all other orbital parameters the same as their current values.*

In Figure 2, we show the variation in the maximum value of orbital eccentricity achieved by the Earth over the one million years of integration time, as a function of Jupiter's orbital eccentricity and semi-major axis. Since the maximum eccentricities achieved by the Earth spanned almost two orders of magnitude, the data is plotted in a logarithmic colour scale, with the most eccentric solutions being displayed in red, and the least eccentric shown in blue. It is immediately apparent that the evolution of Earth's eccentricity varies dramatically as a function of Jupiter's initial orbit. More surprisingly, the plot reveals a great deal of fine structure in the eccentricity evolution of the Earth. Rather than the smooth variation from more stable to less stable orbits that one might expect as Jupiter is moved around, we instead see bands of relatively strong instability and stability running from left to right across the figure. The broad pattern is simple and relatively clear cut – as Jupiter is moved outward, towards the orbit of Saturn, the typical maximum eccentricities experienced by the Earth increase. Similarly, there are more cases where the Earth's eccentricity is driven to large values when Jupiter starts on an eccentric orbit than when it begins on a near circular orbit. However, the fine structure revealed in Fig. 2 shows that, beyond these general, broad brush stroke results, the influence of Jupiter on Earth's orbital eccentricity is actually remarkably complex. There are regions with relatively small excursions for scenarios where Jupiter is both distant from the Sun and relatively eccentric (such as the tongue of blue that extends to the right at an initial eccentricity of ~0.16), and others where the Earth experiences large orbital eccentricities when Jupiter is on a near-circular orbit, close in to the Sun (e.g. the tongue of yellow extending to the left at the bottom of the figure). The cause of this fine structure remains to be tied down, but it seems most likely to be the result of long-term secular effects – and possibly the shifting of multi-body secular resonances as a function of Jupiter's orbital eccentricity (e.g. [40]). Were the variations instead the result of mean-motion resonant[5] interactions between Jupiter and other planets in the system, one would instead expect to see vertical structure in the figure, since mean-motion resonances are purely a function of heliocentric distance. It is also apparent that, although our Solar system (marked by the hollow circle) falls in a region of only moderate variability in eccentricity for the Earth, there are many architectures for the Solar system that would feature significantly smaller excursions in eccentricity for our planet.

---

[5] A mean motion resonance between two bodies occurs when their orbital periods are an integer ratio of one another. The most famous mean motion resonance in our Solar system is that between Neptune and the dwarf planet Pluto. Pluto completes two orbits in the time it takes Neptune to complete three. Even though the orbit of Pluto crosses that of Neptune, their commensurate orbital periods prevent their ever experiencing sufficiently close encounters to disrupt their orbits. Further discussion of resonant orbital behaviour is beyond the scope of this work, but we direct the interested reader to e.g. [41], [42] and [43] for further discussions of orbital resonance and our Solar system.

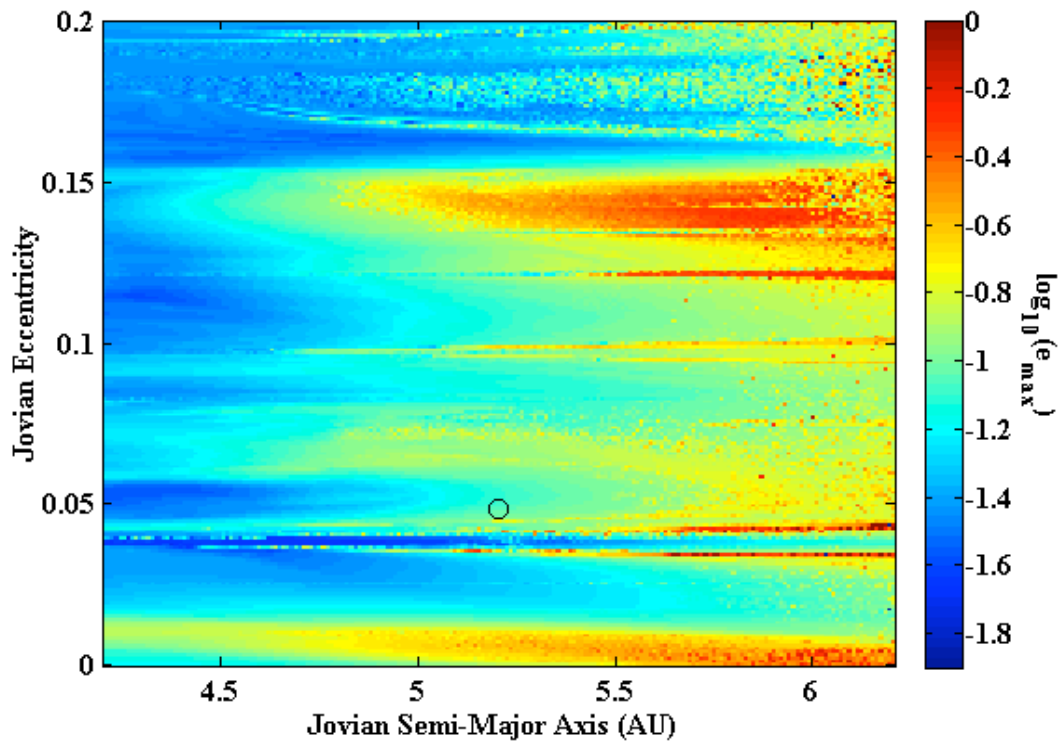

*Fig. 2: The maximum eccentricity of the Earth's orbit obtained over a period of one million years as a function of Jupiter's initial orbital eccentricity and semi-major axis. For each of the 39,601 simulations shown here, the only variables changed in the initial conditions were Jupiter's semi-major axis and eccentricity – the initial orbits of the other planets were held constant across the suite of integrations. The hollow circle shows the location of Jupiter's orbit within our Solar system.*

In Figure 3, we present the variability of the maximum inclination of the Earth's orbit as a function of Jupiter's initial semi-major axis and eccentricity. Here, the variability is much less pronounced than was the case for the Earth's orbital eccentricity. Despite this, a number of similar features are shown, with fingers of increased inclination variability running from left to right in the plot. Once again, a concentration of solutions that feature relatively extreme excursions in the Earth's orbital inclination are concentrated at low Jovian eccentricities, for scenarios where the giant planet is located beyond around 5.5 AU from the Sun.

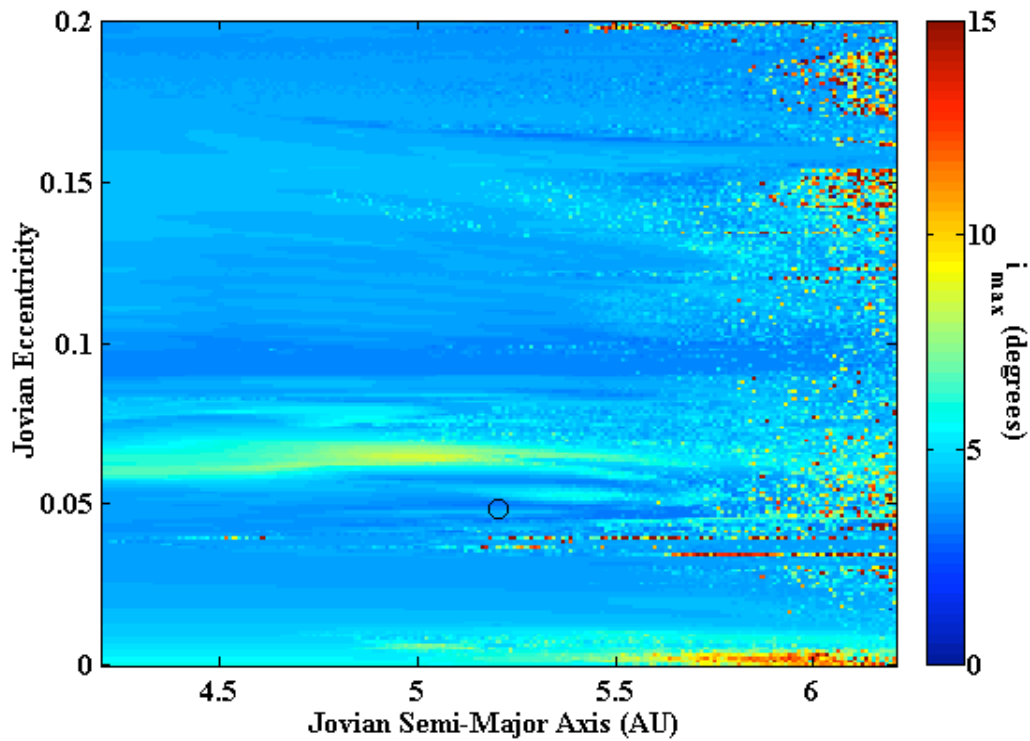

*Fig. 3: The maximum inclination of Earth's orbit over a period of one million years as a function of Jupiter's initial orbital eccentricity and semi-major axis. The hollow circle again marks the location of Jupiter's orbit in our own Solar system.*

Figure 4 shows the manner in which the maximum rate of change of the Earth's orbital semi-major axis varies as a function of Jupiter's eccentricity and orbital radius. In the great majority of cases, the Earth did not experience significant excursions in orbital radius – but although the results here are significantly more noisy than those presented in Figures 2 and 3, similar features are once again visible, with fingers of increased instability stretching from right to left across the figure. It is interesting to compare the results shown in Figures 2 and 4. There are a number of similarities between the two (such as the region of enhanced variability that surrounds the location of our Solar system on three sides). However, although the general structure of the two figures is very similar, it is noticeable that the strength of the different features in different locations varies significantly between the two plots. Compare, for example, the two broad regions of enhanced variability at both high Jovian eccentricity and semi-major axis, at the top right hand corner of both figures. Though the two bands of variability that are visible there display the same sculpting in both figures, the lower of the two results in far more intense variability in the eccentricity of the Earth's orbit, whilst the upper yields far faster variability in the Earth's semi-major axis.

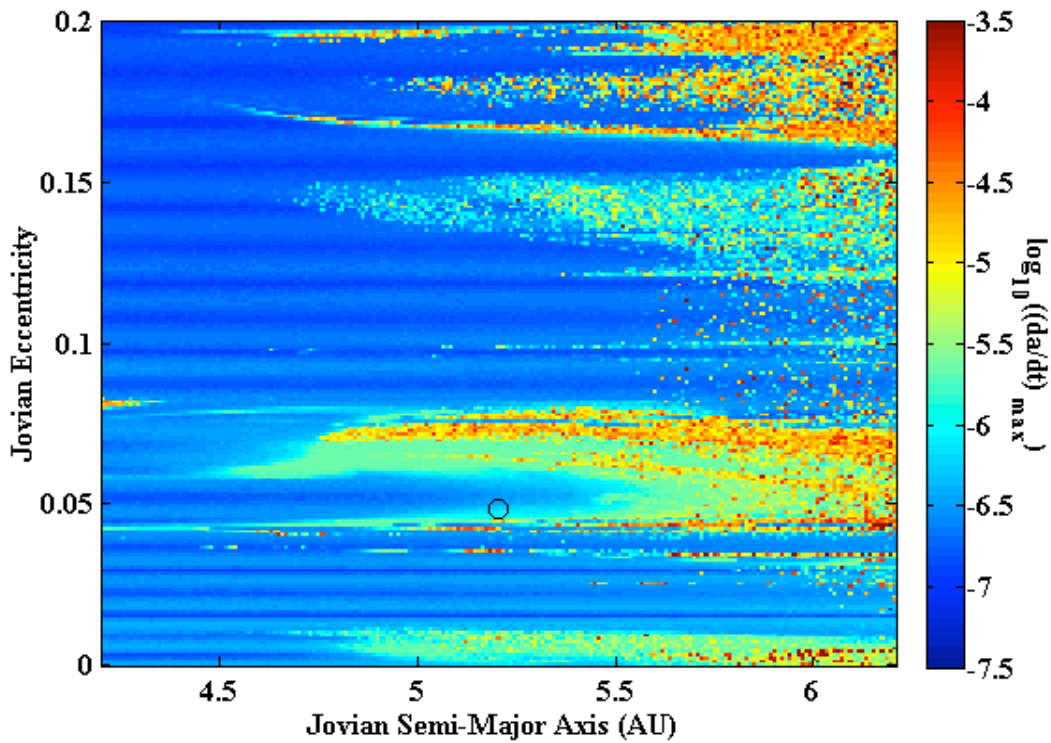

*Fig. 4: The maximum rate of change of the Earth's semi-major axis over a period of one million years, as a function of Jupiter's initial orbital eccentricity and semi-major axis, in units of log$_{10}$ (AU/yr). Variations in the semi-major axis of the Earth were relatively small, but the figure shows how minor variations in Jupiter's initial orbit can cause the rate of that change to vary by up to four orders of magnitude.*

## Conclusions

We have presented the preliminary results of a study that will investigate the influence of the orbit of the giant planet Jupiter on the Milankovitch cycles experienced by the Earth. Our results show that both the amplitude and frequency of the Milankovitch cycles would vary strongly as a function of Jupiter's semi-major axis and eccentricity. Whilst one might expect that those variations would be relatively smooth and easy to predict, we instead find that a great deal of fine structure is present – with Solar systems that differ by only minor changes in Jupiter's semi-major axis or eccentricity displaying significant differences in the strength and periodicity of the Milankovitch cycles.

The next step in this work is to carry out more detail, and longer, simulations of our Solar system. It is possible that our choice to limit our preliminary tests to just one million years of run time will have abridged the longer-period oscillations for some of the more stable systems tested. As a result, we intend to perform simulations that last for an order of magnitude longer over the coming months, using UNSW's *Katana* supercomputing cluster. It is also apparent from our plots (Figures 2 – 4), that there is significant noise around the more unstable solutions tested. We note that a number of our integrations resulted in critically unstable systems, in which one or other of the Solar system's planets were ejected or collided with one another within our 1 million year time frame. Clearly, such unstable systems can result in dramatic variations in the Earth's orbital elements, even when it is not the planet removed in this way – and this is the main reason for the speckle visible towards the highest eccentricities and semi-major axes. We will therefore more than quadruple our resolution in the new suite of runs – testing 159,201 unique architectures for our planetary system, rather than the 39,601 presented in this work. In addition to calculating the variation of the Earth's orbital elements as a function of time, we will also examine the precession of Jupiter's orbit, which will no doubt vary significantly as a function of its initial semi-major axis and eccentricity. Given the precessional frequency of Jupiter's orbit, it will be possible to determine the stability (or instability) of the Earth's obliquity, following the procedure detailed in [44].

Once those simulations are complete, we will collaborate with colleagues at UNSW's Climate Change Research Centre to bring together our dynamical models of the Solar system with simple climate models, in order to assess the degree to which the observed changes in the Earth's orbit across the variety of Jovian orbital architectures studied would affect the climate of the Earth. This will allow us to get a true handle on the importance of planetary architecture as a drive for the climate of exo-Earths, allowing us to build a toolset that will facilitate the study of such planets as and when they are discovered in the coming years.

The long-term goal of this work is to provide a mechanism by which the climatic variability of newly discovered exo-Earths can be estimated, allowing the potential habitability of those worlds to be assessed. Such assessment will form a critical component of the target selection process for the search for life on planets beyond our Solar system. Since the observations that will be required in order to carry out that search will be extremely challenging, it will not be possible to search more than the best few candidates, and so it is imperative that every avenue be explored in order to ensure that we can best select the most promising targets to go forward with the search (e.g. [18][19]).

## Acknowledgments

The work was supported by iVEC through the use of advanced computing resources located at the Murdoch University, in Western Australia.

## References

1. Mayor, M. and Queloz, D., 1995, "A Jupiter-mass companion to a solar-type star", Nature, 378, 355-359

2. Marcy, G. W. and Butler, R. P., 1996, "A Planetary Companion to 70 Virginis", Astrophysical Journal Letters, 464, L147

3. Butler, R. P. and Marcy, G. W., 1996, "A Planet Orbiting 47 Ursae Majoris", Astrophysical Journal Letters, 464, L153

4. Hellier, C., Anderson, D. R., Collier Cameron, A., Gillon, M., Jehin, E., Lendl, M., Maxted, P. F. L., Pepe, F., Pollacco, D., Queloz, D., Ségransan, D., Smalley, B., Smith, A. M. S., Southworth, J., Triaud, A. H. M. J., Udry, S. and West, R. G., 2011, "WASP-43b: the closest-orbiting hot Jupiter", Astronomy and Astrophysics, 535, article id. L7

5. Howard, A. W., Sanchis-Ojeda, R., Marcy, G. W., Johnson, J. A., Winn, J. N., Isaacson, H., Fischer, D. A., Fulton, B. J., Sinukoff, E. and Fortney, J. J., 2013, "A rocky composition for an Earth-sized exoplanets", Nature, early online access: http://www.nature.com/nature/journal/vaop/ncurrent/full/nature12767.html

6. Pepe, F., Collier Cameron, A., Latham, D. W., Molinari, E., Udry, S., Bonomo, A. S., Buchhave, L. A., Charbonneau, D., Cosetino, R., Dressing, C. D., Dumusque, X.,Figueria, P., Fiorenzano, A. F. M., Gettel, S., Harutyunyan, A., Haywood, R. D., Horne, K., Lopez-Morales, M., Lovis, C., Malavolta, L., Mayor, M., Micela, G., Motalebi, F., Nascimbeni, V., Phillips, D., Piotto, G., Pollacco, D., Queloz, D., Rice, K., Sasselov, D., Segransan, D., Sozzetti, A., Szentgyorgyi, A. and Watson, C. A., 2013, "An Earth-sized planet with an Earth-like density", *Nature*, early online access: http://www.nature.com/nature/journal/vaop/ncurrent/full/nature12768.html

7. Doyle, L. R., Carter, J. A., Fabrycky, D. C., Slawson, R. W., Howell, S. B., Winn, J. N., Orosz, J. A., Prcaronsa, A., Welsh, W. F., Quinn, S. N., Latham, D., Torres, G., Buchhave, L. A., Marcy, G. W., Fortney, J. J., Shporer, A., Ford, E. B., Lissauer, J. J., Ragozzine, D., Rucker, M., Batalha, N., Jenkins, J. M., Borucki, W. J., Koch, D., Middour, C. K., Hall, J. R., McCauliff, S., Fanelli, M. N., Quintana, E. V., Holman, M. J., Caldwell, D. A., Still, M., Stefanik, R. P., Brown, W. R., Esquerdo, G. A., Tang, S., Furesz, G., Geary, J. C., Berlind, P., Calkins, M. L., Short, D. R., Steffen, J. H., Sasselov, D., Dunham, E. W., Cochran, W. D., Boss, A., Haas, M. R., Buzasi, D., and Fischer, D., 2011, "Kepler-16: A Transiting Circumbinary Planet", *Science*, 333, 1602

8. Welsh, W. F., Orosz, J. A., Carter, J. A., Fabrycky, D. C., Ford, E. B., Lissauer, J. J., Prvsa, A., Quinn, S. N., Ragozzine, D., Short, D. R., Torres, G., Winn, J. N., Doyle, L. R., Barclay, T., Batalha, N., Bloemen, S., Brugamyer, E., Buchhave, L. A., Caldwell, C., Caldwell, D. A., Christiansen, J. L., Ciardi, D. R., Cochran, W. D., Endl, M., Fortney, J. J., Gautier, III, T. N., Gilliland, R. L., Haas, M. R., Hall, J. R., Holman, M. J., Howard, A. W., Howell, S. B., Isaacson, H., Jenkins, J. M., Klaus, T. C., Latham, D. W., Li, J., Marcy, G. W., Mazeh, T., Quintana, E. V., Robertson, P., Shporer, A., Steffen, J. H., Windmiller, G., Koch, D. G. and Borucki, W. J., 2012, "Transiting circumbinary planets Kepler-34 b and Kepler-35 b" *Nature*, 481, 475-479

9. Bailes, M., Bates, S. D., Bhalerao, V., Bhat, N. D. R., Burgay, M., Burke-Spolaor, S., D'Amico, N., Johnston, S., Keith, M. J., Kramer, M., Kulkarni, S. R., Levin, L., Lyne, A. G., Milia, S., Possenti, A., Spitler, L., Stappers, B. and van Straten, W., 2011, "Transformation of a Star into a Planet in a Millisecond Pulsar Binary", *Science*, 333, 1717

10. Madhusudhan, N., Lee, K. K. M. and Mousis, O., 2012, "A Possible Carbon-rich Interior in Super-Earth 55 Cancri e", *The Astrophysical Journal Letters*, 759, L40

11. O'Toole, S. J., Tinney, C. G., Jones, H. R. A., Butler, R. P., Marcy, G. W., Carter, B. and Bailey, J., 2009, "Selection functions in Doppler planet searches", *Monthly Notices of the Royal Astronomical Society*, 392, 641-654

12. Malhotra, R., 1995, "The Origin of Pluto's Orbit: Implications for the Solar System Beyond Neptune", *Astronomical Journal*, 110, 420

13. Morbidelli, A., Levison, H. F., Tsiganis, K. and Gomes, R., 2005, "Chaotic capture of Jupiter's Trojan asteroids in the early Solar System", *Nature*, 435, 462-465

14. Lykawka, P. S., Horner, J., Jones, B. W. and Mukai, T., 2009, "Origin and dynamical evolution of Neptune Trojans – I. Formation and planetary migration", *Monthly Notices of the Royal Astronomical Society*, 398, 1715-1729

15. Thommes, E. W., Duncan, M. J. and Levison, H. F., 2003, "Oligarchic growth of giant planets", *Icarus*, 49, 195-236

16. Mousis, O., Alibert, Y., Hestroffer, D., Marboeuf, U., Dumas, C., Carry, B., Horner, J. and Selsis, F., 2008, "Origin of volatiles in the main belt", *Monthly Notices of the Royal Astronomical Society*, 383, 1269-1280

17. Fogg, M. J. and Nelson, R. P., 2007, "The effect of type I migration on the formation of terrestrial planets in hot-Jupiter systems", *Astronomy and Astrophysics*, 472, 1003-1015

18. Horner, J. and Jones, B. W., 2010, "Determining habitability: which exoEarths should we search for life?", *International Journal of Astrobiology*, 9, 273-291

19. Horner, J., 2014, "Beyond the Habitable Zone – Dynamics and Habitability", *Life, submitted*

20. Croll, J., 1875, "Climate and time in their geological relations; a theory of secular changes of the earth's climate", Published by Daldy, Tsbister & co., London

21. Köppen, W., Wegener, A., 1924, "Die Klimate der Geologischen Vorzeit", Gebrüder Borntraeger, Berlin.

22. Milankovitch, M., 1941, "Kanon der Erdbestrahlung und seine Andwendung auf das Eiszeiten-problem", R. Serbian Acad., Belgrade

23. Roe, G., 2006, "In defense of Milankovitch", *Geophysical Review* Letters, 33, L24703

24. Hays, J. D., Imbrie, J. and Shackleton, N. J., 1976, "Variations in the Earth's Orbit: Pacemaker of the Ice Ages", *Science*, 194, 1121-1132

25. Kopp, R. E., Kirschvin,, J. L., Hilburn, I. A. and Nash, C. Z., 2005, "The Paleoproterozoic snowball Earth: A Climate disaster triggered by the evolution of oxygenic photosynthesis", *Proceedings of the National Academy of Sciences of the United States of America*, 102, 11131-11136

26. Preiss, W. V., Gostin, V. A., McKirdy, D. M., Ashley, P. M., Williams, G. E. and Schmidt, P. W., 2011, "Chapter 69 The glacial succession of Sturtian age in South Australia: the Yudnamutana Subgroup", *Geological Society, London, Memoirs*, 36, 701-712

27. Shields, G. A., 2008, "Palaeoclimate: Marinoan meltdown", *Nature Geoscience*, 1, 351-353

28. Finnegan, S., Bergmann, K., Eiler, J. M., Jones, D. S., Fike, D. A., Eisenman, I., Hughes, N. C., Tripati, A. K. and Fischer, W. W., 2011, "The magnitude and duration of Late Ordovician-Early Silurian glaciation", *Science*, 331, 903-906

29. Williams, G. E., 1991, "Milankovitch-band cyclicity in bedded halite deposits contemporaneous with Late Ordovician-Early Silurian glaciation, Canning Basin, Western Australia", *Earth and Planetary Science Letters*, 103, 143-155

30. Strom, R. G., Sprague, A. L., "Exploring Mercury: the iron planet", eds. Strom, R. G., Sprague, A. L., Springer-Praxis Books in Astronomy and Space Sciences, London (UK): Springer, Chichester (UK): Praxis Publishing. ISBN 1-85233-731-1, 2003

31. Correia, A. C. M., Laskar, J., 2009, "Mercury's capture into the 3/2 spin–orbit resonance including the effect of core–mantle friction", *Icarus*, 201, 1-11

32. Chambers, J. E., 1999, "A hybrid symplectic integrator that permits close encounters between massive bodies", *Monthly Notices of the Royal Astronomical Society*, Volume 304, pp. 793 – 799

33. Horner, J., Lykawka, P. S., Bannister, M. T., Francis, P., 2012, "2008 LC18: a potentially unstable Neptune Trojan", *Monthly notices of the Royal Astronomical society*, 422, 2145-2151

34. Robertson, P., Endl, M., Cochran, W. D., MacQueen, P. J., Wittenmyer, R. A., Horner, J., Brugamyer, E. J., Simon, A. E., Barnes, S. I., Caldwell, C., "The McDonald Observatory Planet Search: New Long-period Giant Planets and Two Interacting Jupiters in the HD 155358 System", *Astrophysical Journal*, 739, article id. 39

35. Horner, J., Jones, B. W., 2009, "Jupiter – friend or foe? II: the Centaurs", *International Journal of Astrobiology*, 8, 75-80

36. Horner, J., Lykawka, P. S., 2010, "2001 QR322: a dynamically unstable Neptune Trojan?", *Monthly Notices of the Royal Astronomical Society*, 405, 49-56

37. Horner, J., Wittenmyer, R. A., Hinse, T. C., Tinney, C. G., 2012, "A detailed investigation of the proposed NN Serpentis planetary system", *Monthly Notices of the Royal Astronomical Society*, 425, 749-756

38. Wittenmyer, R. A., Endl, M., Cochran, W. F., Levison, H. F. and Henry, G. W., 2009, "A Search for Multi-Planet Systems Using the Hobby-Eberly Telescope", *The Astrophysical Journal Supplement*, 182, 97-119

39. Moutou, C., Hébrard, G., Bouchy, F., Eggenberger, A., Boisse, I., Bonfils, X., Gravallon, D., Ehrenreich, D., Forveille, T., Delfosse, X., Desort, M., Lagrange, A.-M., Lovis, C., Mayor, M., Pepe, F., Perrier, C., Pont, F., Queloz, D., Santos, N. C., Ségransan, D., Udry, S. and Vidal-Madjar, A., 2009, "Photometric and spectroscopic detection of the primary transit of the 111-day-period planet HD 80 606 b", *Astronomy and Astrophysics*, 498, L5-L8

40. Minton, D. A., Malhotra, R., 2011, "Secular Resonance Sweeping of the Main Asteroid Belt During Planet Migration", *The Astrophysical Journal*, 732, article id. 53

41. Murray, C. D., Dermott, S. F., "Solar system dynamics", ed. Murray, C. D., Cambridge University Press, 1999

42. Horner, J., Jones, B. W., 2008, "Jupiter friend or foe? I: The asteroids", *International Journal of Astrobiology*, 7, 251-261

43. Lykawka, P. S., 2012, "Trans-Neptunian Objects as Natural Probes to the Unknown Solar System", *Monographs on Environment, Earth and Planets*, 1, 121-186

44. Waltham, D., 2006, "The large-moon hypothesis: can it be tested?", *International Journal of Astrobiology*, 5, 327-331

45. Speigel, D. S., Raymond, S. N., Dressing, C. D., Scharf, C. A. and Mitchell, J. L., 2010, "Generalized Milankovitch Cycles and Long-Term Climatic Habitability", *The Astrophysical Journal*, 721, 1308-1318